\documentclass[sigconf]{acmart}

\AtBeginDocument{%
  }

\copyrightyear{2025}
\acmYear{2025}
\acmConference[IUI '25]{30th International Conference on Intelligent User Interfaces}{March 24--27, 2025}{Cagliari, Italy}
\acmBooktitle{30th International Conference on Intelligent User Interfaces (IUI '25), March 24--27, 2025, Cagliari, Italy}\acmDOI{10.1145/3708359.3712108}
\acmISBN{979-8-4007-1306-4/25/03}

\usepackage{wrapfig}  
\usepackage{soul}  
\usepackage{fontawesome}  

\newcommand{\reviewAdd}{\textcolor{black}}

\newcommand{\purpleNode}{\textcolor[RGB]{195, 158, 255}}
\newcommand{\grayNode}{\textcolor[RGB]{136, 136, 136}}
\newcommand{\lorangeNode}{\textcolor[RGB]{255, 196, 108}}
\newcommand{\dorangeNode}{\textcolor[RGB]{255, 159, 108}}
\newcommand{\pauseArea}{\textcolor[RGB]{0, 174, 236}}
\newcommand{\speedLine}{\textcolor[RGB]{255, 152, 151}}

\begin{document}

\title[TSConnect]{TSConnect: An Enhanced MOOC Platform for Bridging Communication Gaps Between Instructors and Students in Light of the Curse of Knowledge}

 \author{Qianyu Liu}
 \email{liuqy@shanghaitech.edu.cn}
 \orcid{0009-0006-0212-1318}
 \affiliation{%
   \institution{School of Information Science and Technology, ShanghaiTech University}
   \city{Shanghai}
   \country{China}}

  \author{Xinran Li}
  \email{adelineli@vt.edu}
  \orcid{0009-0006-9503-129X}
  \affiliation{%
    \institution{Department of Computer Science, Virginia Tech}
    \city{Blacksburg, Virginia}
    \country{United States}
  }

  \author{Xiaocong Du}
  \email{duxc2023@shanghaitech.edu.cn}
  \orcid{0009-0008-1251-5171}
  \affiliation{%
    \institution{School of Information Science and Technology, ShanghaiTech University}
    \city{Shanghai}
    \country{China}}

 \author{Quan Li}
 \authornote{Corresponding Author.}
 \email{liquan@shanghaitech.edu.cn}
 \orcid{0000-0003-2249-0728}
 \affiliation{%
   \institution{School of Information Science and Technology, ShanghaiTech University}
   \city{Shanghai}
   \country{China}
 }

\renewcommand{\shortauthors}{Liu et al.}

\begin{abstract}
    Knowledge dissemination in educational settings is profoundly influenced by the curse of knowledge, a cognitive bias that causes experts to underestimate the challenges faced by learners due to their own in-depth understanding of the subject. This bias can hinder effective knowledge transfer and pedagogical effectiveness, and may be exacerbated by inadequate instructor-student communication. To encourage more effective feedback and promote empathy, we introduce \textit{TSConnect}, a bias-aware, adaptable interactive MOOC (Massive Open Online Course) learning system, informed by a need-finding survey involving $129$ students and $6$ instructors. \textit{TSConnect} integrates instructors, students, and Artificial Intelligence (AI) into a cohesive platform, facilitating diverse and targeted communication channels while addressing previously overlooked information needs. A notable feature is its dynamic knowledge graph, which enhances learning support and fosters a more interconnected educational experience. We conducted a between-subjects user study with $30$ students comparing \textit{TSConnect} to a baseline system. Results indicate that \textit{TSConnect} significantly encourages students to provide more feedback to instructors. Additionally, interviews with $4$ instructors reveal insights into how they interpret and respond to this feedback, potentially leading to improvements in teaching strategies and the development of broader pedagogical skills.
\end{abstract}




\maketitle

\section{Introduction}
\par Education serves as a cornerstone for personal growth, societal progress, and economic prosperity~\cite{Hanushek:2012:Better}. In this context, instructors and educators wield significant influence over the acquisition of knowledge by students and novices, thereby shaping the evolution of various scientific disciplines~\cite{Wiemann:2007:curse,Shindler:2009:Transformative}. However, discussions about the shortcomings of educational systems often spotlight a prevalent cognitive bias known as \textbf{the curse of knowledge}, particularly pronounced among instructors teaching engineering and science subjects at the tertiary level~\cite{Wiemann:2007:curse,Fisher:2016:TheCurse,Ambrose:2010:Learning}. This bias arises when instructors unintentionally overlook the unfamiliar and uncertain experiences encountered by learners when grappling with new concepts~\cite{Heath:2007:Made,Learn:2000:Brain,Zwick:1995:OnthePractical}. Their deep expertise and profound subject understanding may hinder effective knowledge transmission, leading instructors to underestimate the challenges faced by students in comprehending new material~\cite{Wiemann:2007:curse,Ambrose:2010:Learning}. This underscores the importance of relying not solely on faculty opinions but also on validated student feedback and assessment methods to enhance learning outcomes~\cite{Gray:1992:National,Pipia:2022:Curse}.

\par In the preparation phase, instructors meticulously organize the material to be covered in upcoming classes, drawing from the prescribed syllabus~\cite{Parkes:2002:Purposes}. In addition to introducing new topics, they often opt to review fundamental or prerequisite concepts, drawing upon their own teaching acumen and insights into student needs~\cite{Stockard:2018:Effectiveness}. Throughout lectures, instructors dynamically adapt their delivery and explanations, integrating real-time feedback from students~\cite{Munna:2021:Teaching}. This process involves striking a delicate balance between catering to the comprehension levels of the majority of students and meeting the standard requirements of instruction, as revealed through interviews with instructors (see \autoref{sec:formativeStudy}). Both online and in-person modalities are applicable for this approach, albeit with slightly distinct feedback mechanisms.

\par Despite the pivotal role of instructors in education, traditional instructor-centred approaches often fall short in meeting the diverse needs and preferences of students~\cite{Shindler:2009:Transformative}. The transmission of new knowledge faces two significant challenges. First, \textbf{in the preparation phase, instructors frequently struggle to accurately assess students' levels of prerequisite knowledge}, necessitating continual adjustment during lectures. Given the diverse educational backgrounds and learning paths of students, accurately gauging their knowledge reserves proves challenging~\cite{Pipia:2022:Curse}. While instructors possess a comprehensive understanding of the interconnectedness and context of knowledge within their field, students typically have only been exposed to a fraction of this domain~\cite{Nathan:2003:Expert}. Consequently, instructors may overlook gaps in students' prerequisite knowledge, exacerbated by the tendency for students to forget previously learned material to varying degrees~\cite{Ebbinghaus:2013:Memory}. This oversight may result in the introduction of more complex concepts before students have mastered fundamental knowledge, impeding systematic learning and potentially undermining student motivation. Second, \textbf{during lectures, instructors may struggle to accurately gauge the learning progress of their students}. For example, in interactive classroom settings, students may not consistently provide instructors with effective and genuine feedback, leading to misunderstandings about classroom dynamics. Students may have difficulty accurately assessing their own comprehension and articulating the root of their difficulties, often hesitating to ask questions in class. These issues are further magnified in online teaching environments~\cite{Ma:2022:Glancee}. Moreover, subsequent assessment methods, such as assignments and exams, frequently struggle to offer specific and timely feedback on classroom performance.

\par Technology-enhanced learning (TEL)~\cite{Sen:2019:TEL} approaches, integrated with machine learning techniques, are garnering increased recognition for addressing challenges from both instructors' and students' perspectives~\cite{Ashok:2022:ASystematic,Kuka:2022:Teaching}. For instructors' convenience, some studies have focused on automatically detecting students' learning statuses and aggregated feedback during classes~\cite{Liu:2018:Towards,Chiu:2021:ABayesian,Rivera:2013:Live,Chamillard:2011:Using,Ma:2022:Glancee}. Others have proposed intelligent tutoring agents to support personalized learning before or after class, offering suggestions for further instructions~\cite{Holstein:2017:Intelligent,Benotti:2014:Engaging,Holmes:2023:Guidance,Diwanji:2018:Enhance}. While these efforts streamline teaching activities and provide recommendations, they primarily target existing instructional problems rather than enhancing teaching ability and fostering empathy towards students. In particular, \textbf{current TEL approaches overlook assisting instructors in raising awareness about the curse of knowledge.} Although educational researchers have summarized various strategies to mitigate this bias~\cite{Froyd:2008:Faculty,Ambrose:2010:Learning,Heath:2007:Made,Pipia:2022:Curse}, practical application often proves challenging, as educators are encouraged to refine their approaches by closely observing students' cognitive processes in real-world contexts~\cite{Wiemann:2007:curse}. In other words, \textbf{theoretical training aimed at bias awareness may lose efficacy in actual teaching scenarios}~\cite{Carter:2020:Developing}. For students, many learning recommendation systems have been introduced to generate personalized learning paths, either to expand existing knowledge~\cite{Xia:2019:PeerLens,Murayama:2023:IKnowde} or to identify and bridge knowledge gaps in specific subject areas~\cite{Okubo:2023:Adaptive,Zheng:2020:Fill,Bauman:2018:Recommending}. However, \textbf{limited consideration has been given to identifying prerequisite gaps that hinder the acquisition of new content}, which directly impedes learning in a more systematic manner. Furthermore, most studies \textbf{have neglected cognition gaps in student-instructor communication}, where students often struggle to articulate their questions and instructors face challenges in comprehension, particularly aligning with the teaching material.

\par This study centers on online teaching, which, despite its limitations such as the absence of non-verbal cues, presents significant advantages for learning data collection and is well-suited for TEL applications. By utilizing existing course videos and online platforms, instructors can gain insights into students' needs and preferences, tailoring teaching content accordingly through the analysis of student interactions and feedback. Moreover, there is potential to enrich existing videos to offer students a more structured and contextually relevant learning experience. Consequently, our aim is to establish a loop involving instructors, students, and artificial intelligence (AI) to address biases effectively.
To explore instructors' and students' actual information needs and preferences, as suggested by prior literature~\cite{Pipia:2022:Curse}, and to assess the feasibility of integrating such information into a comprehensive education recommendation system, we aim to address two primary research questions: \textbf{RQ1: How do instructors and students perceive and cope with instructors' curse of knowledge?} and \textbf{RQ2: What methods are deemed acceptable for mitigating bias and raising awareness?} To address \textbf{RQ1}, we conducted a survey involving $129$ students from various academic backgrounds and degrees, complemented by expert interviews with $6$ instructors across different disciplines at a local university. Analysis of the survey and interview findings revealed that the lack of spontaneous student feedback contributes to the persistence of the curse of knowledge in educational settings. Based on this feedback, we identified three design requirements for each user end for the system to address \textbf{RQ2}. Subsequently, we developed an adaptable online MOOC (Massive Open Online Course) learning system named \textit{TSConnect}. This system collects diverse leaning and feedback data to help instructors gauge students' knowledge levels and monitor their learning progress. Additionally, students can access guidance on prerequisite knowledge required for their current learning process. In the frontend for students, we provide an interactive dynamic knowledge graph alongside lecture videos, serving as a novel data collection interface and aiding systematic learning. In the frontend for instructors, we offer a \textit{VideoData View} and \textit{Network View} for retrospective review and analysis, assisting instructors in pinpointing instances where the curse of knowledge may arise that contribute to learning challenges.

\par Through the proposed research prototype, we further explored the following research questions: \textbf{RQ3: What is the usability and effectiveness of the support system?} \textbf{RQ4: How do students(RQ4-a) and instructors(RQ4-b) perceive the support system?} and \textbf{RQ5: What impact does the support system have on current teaching and learning practices?} To address these questions, we conducted a between-subjects user study involving $30$ students hailing from a local university. Students engaged with multiple course videos under two different conditions: one with the proposed \textit{TSConnect} and the other as a baseline condition where students solely viewed videos and sent textual comments, with their interaction data collected for later analysis. By administering post-task surveys to student participants and comparing their feedback data logs, we ascertained that \textit{TSConnect} effectively motivates more frequent and comprehensible feedback, as evidenced by survey results. Additionally, we conducted expert interviews with instructor participants, probing their understanding of feedback data and the impact on their current and future pedagogical practice. This work makes the following contributions:
\begin{itemize}
\item We conducted a survey with $129$ students to assess their perceptions of biased teaching and interviewed $6$ instructors to understand their awareness of the curse of knowledge and their needs for understanding students.
\item We developed \textit{TSConnect}, an online platform that integrates dynamic knowledge graph algorithms to enhance the learning experience and facilitates instructors' acquisition of student feedback.
\item We performed a between-subjects user study to evaluate the usability, effectiveness, and user interaction patterns of \textit{TSConnect}, and investigate its potential implications for educational practices.
\end{itemize}


\section{Related Work}

\subsection{The Curse of Knowledge}
\par Extensive research has delved into the phenomenon known as the Curse of Knowledge, identifying it as a cognitive bias prevalent across various domains~\cite{Camerer:1989:TheCurse,Sadler:2013:TheInfluence,Xiong:2020:TheCurse}. Within the realm of communication, individuals often subconsciously assume that their counterparts possess the necessary background knowledge to fully grasp their message~\cite{Zwick:1995:OnthePractical,Learn:2000:Brain}. This tendency is particularly pronounced in educational contexts~\cite{Froyd:2008:Faculty}, where the curse of knowledge can significantly hinder effective teaching and learning~\cite{Wiemann:2007:curse}. Heath et al.~\cite{Heath:2007:Made} have defined this phenomenon as the disconnect between educators, who possess knowledge, and learners, who lack it. Specifically, instructors frequently overestimate their students' familiarity with the subject matter being taught~\cite{Sadler:2013:TheInfluence,Pipia:2022:Curse}. Previous research has attributed this discrepancy to instructors' heavy reliance on their own expertise~\cite{Wiemann:2007:curse,Sadler:2013:TheInfluence}, insufficient consideration of students' perspectives~\cite{Wiemann:2007:curse,Ambrose:2010:Learning}, or a lack of diagnostic cues regarding students' existing knowledge~\cite{Pipia:2022:Curse,Tullis:2022:TheCurse}.

\par To overcome this curse, Heath et al.~\cite{Heath:2007:Made} outlined six key factors to consider. Expanding upon this research, Froyd et al.~\cite{Froyd:2008:Faculty} developed four strategies aimed at increasing awareness of the curse of knowledge bias and supporting faculty professional development. Ambrose et al.~\cite{Ambrose:2010:Learning} proposed three components to mitigate the curse and identified seven evidence-based principles for enhancing effective learning. Similarly, Pipia et al.~\cite{Pipia:2022:Curse} conducted a qualitative study involving students and instructors to gather insights into educational processes and the operationalization of these seven principles in classroom settings. While physics instructors have access to a wealth of educational research providing insights into students' cognitive processes and common challenges~\cite{McDermott:1999:Resource}, these resources may be insufficient and susceptible to inertia.

\par This study aims to assist instructors in promptly recognizing students' confusion and uncertainty, thereby facilitating improvements in teaching methodologies. Drawing inspiration from theoretical research~\cite{Pipia:2022:Curse}, we address the educational dilemma where instructors may lack awareness of students' prior knowledge and requirements, overlooking their actual capabilities and the need for further clarification when introducing new concepts. To achieve this objective, we advocate for the implementation of a human-machine collaboration approach, aimed at strengthening the connection between students and educators.

\subsection{Technology-Enhanced Learning and Educational Recommendation Systems}
\par Technology-enhanced learning (TEL) includes a wide array of computer-based technologies aimed at facilitating learning~\cite{Sen:2019:TEL}. In line with our research objectives, we narrow our focus to relevant literature on educational recommendation techniques designed to support learning and teaching activities.

\par In conventional settings, students typically need to manually sift through predefined syllabi to identify relevant learning materials, whereas TEL can leverage machine learning techniques to recommend supplementary learning materials from both internal sources (e.g., lecture materials~\cite{Yang:2021:Using}) and external sources (e.g., online articles and videos~\cite{Zhao:2018:Flexible}). Moreover, prior research has demonstrated the potential to design personalized learning pathways for learners. According to Adomavicius and Tuzhilin~\cite{Adomavicius:2005:Toward}, recommendation systems fall into three primary categories: Content-based systems recommend items based on the relationships between knowledge components (e.g., as seen in the work of Murayama et al.~\cite{Murayama:2023:IKnowde}). Collaborative Filtering systems recommend items based on the historical preferences and profiles of similar individuals (e.g., demonstrated by Rafaeli et al.~\cite{Rafaeli:2005:Social}). Hybrid approaches integrate both collaborative and content-based methods (e.g., as shown in the research of Salehi et al.~\cite{Salehi:2013:Hybrid}). Additionally, contextual information such as learner feedback can enhance the learning process~\cite{Derntl:2005:Modeling}. This feedback can be gathered explicitly through methods like questionnaires~\cite{Murayama:2023:IKnowde} or implicitly through measures such as time spent on tasks and click history~\cite{Xia:2019:PeerLens}.

\par Moreover, various recommendation techniques cater to instructors' needs. For instance, Liu et al.~\cite{Liu:2018:Towards} proposed a smart learning recommendation system that utilizes sensor data to suggest effective learning activities in the classroom based on students' current learning states. Ma et al.~\cite{Ma:2022:Glancee} integrated adaptable monitoring and retrospective interfaces with computer vision algorithms to infer students' remote learning status for instructors. In the context of flipped classrooms, AI chatbots~\cite{Diwanji:2018:Enhance} can engage in conversations based on subject matter, interact with students as tutors, and provide teaching strategies and tips for instructors preparing classroom materials. Unlike these approaches, which directly aid instructors in identifying and resolving issues, our objective is to raise instructors' awareness of the curse of knowledge and assist in fostering a student-centered teaching approach.

\par While the aforementioned work can assist both instructors and learners by providing recommendations for subsequent activities or suggesting alternative options, it is also imperative to address the knowledge gap in the subject matter itself. To support after-class knowledge review, Bauman et al.~\cite{Bauman:2018:Recommending} introduced a methodology for identifying unmastered knowledge and recommending remedial learning materials to improve performance in final exams. Okubo et al.~\cite{Okubo:2023:Adaptive} presented a personalized review system that recommends materials tailored to the learner's level of understanding. In contrast to post-class methods, Zheng et al.~\cite{Zheng:2020:Fill} identified learning shortfalls at an early stage by tracking in-class emotions. Beyond addressing post-learning mastery gaps is crucial, identifying prerequisite knowledge gaps is equally essential for sustainable learning. Therefore, we propose a novel approach to derive a past-oriented learning recommendation that emphasizes prerequisite knowledge.

\subsection{Teacher Education and Teaching Skills}
\par ``\textit{Skillful teachers are made, not born}''~\cite{Saphier:1997:Skillful}. Becoming an excellent educator entails not only the acquisition of a broad knowledge base but also the proficiency in conveying knowledge to students in a clear and systematic manner. In the $21^{st}$ century, essential skills like critical thinking have surpassed rote memorization as the primary focus of education~\cite{Dede:2010:Comparing}. The global adoption of Learner-Centred Pedagogy (LCP)~\cite{Schweisfurth:2013:Learner}, which emphasizes understanding and addressing the unique needs and perspectives of each student, has heightened the expectations placed on instructors~\cite{Darling:200:Constructing}. Teacher education is instrumental in equipping educators with the skills necessary to effectively apply LCP principles. It is not sufficient to merely adopt the outward forms of LCP, such as questioning techniques; instructors must fully integrate its substance into their teaching practices~\cite{Brodie:2002:Forms}. Numerous publications within the education domain provide instructional guidance for instructors~\cite{Saphier:1997:Skillful,Almazroa:2023:Teaching,Banner:2017:Elements}. These resources are particularly beneficial for pre-service instructors, providing them with experiential knowledge that extends beyond their personal teaching experiences.

\par The existing literature on instructors skill development includes a variety of interventions~\cite{Beriswill:2016:Professional}, tools~\cite{Fenton:2017:Recommendations}, and frameworks~\cite{Butler:2017:Different}, along with methodologies such as peer observation~\cite{Keiler:2023:Supporting} and self-assessment~\cite{Kim:2019:Improving}. Reflective practice is highlighted as a pivotal element within instructors education, where detailed and specific feedback is essential for fostering sustained and substantive improvements through in-depth analysis and introspection~\cite{Rodgers:2002:Defining,Rivers:2013:Improving}. Recent studies also suggested that large language models (LLMs) could enhance instructors' reflective capacities and encourage innovative practices~\cite{Whalen:2023:Chatgpt}. However, the literature cautions against enforced reflection and rote thinking, which may fail to produce genuine behavioral changes in instructors and could even introduce social desirability bias~\cite{Hobbs:2007Faking}.

\par Reflective practice requires continuous and timely feedback. While peers and third-party expert observations offer valuable objectivity, they could be costly and demand extensive preparatory training, which poses challenges in resource-constrained regions~\cite{Kim:2019:Improving}. Our work aims to enrich existing MOOC platforms by incorporating more granular analyses of student learning behaviors and feedback. The interactive visualizations we provide are designed to encourage instructors to engage in deep reflection and introspection. Unlike previous studies~\cite{Shi:2015:VisMOOC}, our approach extends beyond the examination of video clickstream data by integrating student feedback on key concepts within the videos, offering a more comprehensive and analytical perspective.


\section{Formative Study} \label{sec:formativeStudy} 
\par This study deals with the curse of knowledge bias in teaching by TEL technologies, aiming to enhance the teaching effectiveness and foster greater alignment between instructors and students. We investigated \textbf{RQ1} and \textbf{RQ2} through student surveys and instructor interviews, informing our system design.

\subsection{Survey Study of Students}
\subsubsection{Survey Protocol} 
\par Based on the findings from \cite{Pipia:2022:Curse} and informal discussions with some students, we crafted a survey to collect student's experiences with online classes. The survey covered demographic information, learning challenges, communication patterns with instructors, and attitudes toward learning data analytics. Following IRB approval, we distributed the survey via social media to participants with minimum high school education. Responses were excluded if incomplete or completed under 50 seconds.

\subsubsection{Respondents}
\par We received 129 valid responses from students ($65$ male, $60$ female, and $4$ who preferred not to disclose). The respondents included $17$ high school students, $72$ undergraduates, $35$ master students, and $5$ Ph.D. students. Excluding the high school participants, the respondents represented a wide range of grades and majors, including science, medicine, engineering, business, humanity, and other fields. All students had prior experience with online learning.

\subsection{Semi-structured Interview of instructors}
\subsubsection{Interview Protocol}
\par As detailed in \autoref{tab:fmtInterviewScript}, instructor interview explored participants' class preparation methods. Drawing on student survey results, discussions focused on scenarios related to the curse of knowledge, as well as their coping strategies and specific requirements for TEL tools. We employed Braun and Clarke's six-phase thematic analysis framework to analyze the interview transcripts. Initial coding by one author underwent team review for completeness. Two authors then iteratively refined themes until reaching consensus on key findings.

\subsubsection{Participants}
\par A pilot discussion with an extra instructor informed the development and refinement of the interview protocol. Subsequently, $6$ instructors (I1$\sim$6, $3$ males, $3$ females) participated in formal interviews, who comprised two novice instructors ($M = 4$ years experience) and four experienced instructors ($M = 26.8$ years). As shown in \autoref{tab:fmtinterviewDemo}, participants represented diverse disciplines and institutions. All participants had experience using online educational platforms or tools due to the impact of Covid-19.

\begin{table}[h]
\centering
\begin{tabular}{cccc}
\toprule
ID & Gender/Duration & Instructor Type   & Major              \\ \hline
I1 & Male/27         & high school       & Chemistry          \\
I2 & Female/30       & high school       & Geography          \\
I3 & Male/4          & higher education    & Mathematics      \\
I4 & Female/30       & higher education    & Machine Learning \\
I5 & Male/4          & higher education    & Computer Science \\
I6 & Female/20       & higher education    & Tourism          \\ \bottomrule
\end{tabular}
\caption{Demographic information of interview instructors. Duration denotes the number of years a participant has taught as an instructor. An instructor of higher education implies teaching personnel affiliated with a university or a similar tertiary-level educational establishment.}
\label{tab:fmtinterviewDemo}
\end{table}
\begin{table*}[h]
\begin{tabular}{ll}
\toprule
Category            & Question                                         \\ 
                    \hline
Demographic         & What is your major area of specialty and what courses do you typically instruct?               \\
                    & How long have you been in the teaching profession?                                             \\
                    \hline
                    & What is your overall process for preparing a course and an individual lessons respectively?     \\
                    & How do you design and structure your lecture content?                                           \\
Procedures          & How do you gauge students' prior knowledge and their understanding of new concepts?             \\
                    & How do you get and utilize students' learning feedback?                                         \\
                    & How do you balance your teaching goals and students learning?                                   \\
                    \hline
                    & Have you ever ignore students' basic knowledge levels when preparing lessons?                   \\
Teaching issues \&  & Have you ever misjudged students' grasp of a certain part of the lesson content? \\
potential solutions & Have you ever faced challenges in understanding student feedback?                                \\
                    & What unique challenges exist of online environment, excluding hardware-related issues?          \\
                    \hline
Feedback data       & How do/will you utilize interaction data of MOOC videos to help you solve the teaching issues?  \\
                    & What type of feedback data can better help you to adjust your learning?                         \\
                    \hline
Expectation         & What functions do you want to add or improve to the current MOOC system?                        \\ 
                    \bottomrule
\end{tabular}
\caption{Interview with instructors.}
\label{tab:fmtInterviewScript}
\end{table*}

\subsection{Findings} \label{sec:formativeFinding}
\par This section presents six key findings from surveys and interviews on the curse of knowledge in the current teaching process. Building on the foundational insights of \cite{Pipia:2022:Curse}, our study offers a deeper exploration into the persistent nature of this bias, even as both instructors and students are increasingly aware of its impact.

\begin{table*}[h]
\centering
\begin{tabular}{llcccccc}
\toprule
\multicolumn{8}{r}{\textbf{Do you struggle to comprehend new knowledge and maintaining pace with the curriculum progression?}} \\
& & Never& Seldom& Sometimes& Often& \multicolumn{2}{c}{Always}\\ \hline
 \textbf{Are you willing to} & Strongly Disinclined$^{-}$& & 1& 0& 1& \multicolumn{2}{c}{0}\\
 \textbf{provide feedback}& Disinclined$^{-}$& & 11& 9& 6& \multicolumn{2}{c}{3}\\
  \textbf{to your instructor} & Neutral& 10& 8& 17& 4& \multicolumn{2}{c}{0}\\
 \textbf{regarding your} & Inclined$^{+}$& & 18& 19& 7& \multicolumn{2}{c}{2}\\
 \textbf{difficulties?}& Strongly Inclined$^{+}$& & 8& 3& 2& \multicolumn{2}{c}{0}\\ \hline
 \multicolumn{3}{l}{\textbf{Student difficulties in comprehending}}& \multicolumn{4}{l}{\textbf{Student challenges in providing feedback}}\\ \hline
 \multicolumn{2}{l}{Rapid pace of instruction}& 57/129 \reviewAdd{(44.2\%)}& & & willing&unwilling\\
 \multicolumn{2}{l}{Incomprehensible instructional logic}& 28/129 \reviewAdd{(21.7\%)}& \multicolumn{2}{l}{Feedback mechanism deficiency}& 37/88 \reviewAdd{(42.0\%)}&24/31 \reviewAdd{(77.4\%)}
\\
 \multicolumn{2}{l}{Unawareness of teaching plan}& 26/129 \reviewAdd{(20.2\%)}& \multicolumn{2}{c}{Lack of instructor responsiveness}& 15/88 \reviewAdd{(17.0\%)}&3/31 \reviewAdd{(9.7\%)}
\\
 \multicolumn{2}{l}{Insufficient domain knowledge}& 65/129 \reviewAdd{(50.4\%)}& \multicolumn{2}{c}{Inefficacious instructor's solution}& 18/88 \reviewAdd{(20.4\%)}&3/31 \reviewAdd{(9.7\%)}
\\
 \multicolumn{2}{l}{Insufficient prerequisite knowledge}& 44/129 \reviewAdd{(34.1\%)}& \multicolumn{2}{l}{Self-diagnosis difficulty}& 42/88 \reviewAdd{(47.7\%)}&11/31 \reviewAdd{(35.5\%)}
\\
 \multicolumn{2}{l}{Perceived weak comprehension abilities}& 30/129 \reviewAdd{(23.3\%)}& \multicolumn{2}{l}{No Learning Impediments}& 20/88 \reviewAdd{(22.7\%)}&3/31 \reviewAdd{(9.7\%)}
\\
 \multicolumn{2}{l}{Forgetting previously acquired knowledge}& 42/129 \reviewAdd{(32.6\%)}& & & &\\ \bottomrule
 \end{tabular}
\caption{A total of 129 valid responses were obtained in the survey study of students.}
\label{tab:fmtSurveyResult}
\end{table*}

\subsubsection{[\textbf{Finding 1}] \textbf{The Necessity of instructors' proactive assessment of learning status}}
\par Survey results (as shown in \autoref{tab:fmtSurveyResult}) show that students' average self-assessment of learning effort is $3.29$ (SD=$0.92$) on a 5-point scale, with $1/3$ frequently frustrated. Over $1/2$ struggle to keep up, and a quarter hesitate to communicate challenges. More than $1/2$ feel a mismatch between comprehension abilities and instruction pace. Interview analysis reveals that despite instructors' encouragement, only a subset of students proactively interact, leaving instructors with limited, potentially biased feedback. Instructors often rely on observing students' expressions and use questioning and quizzes to refine their teaching strategies when necessary. However, this observation can be vague, as I5 expressed: ``\textit{When I see students bowing their heads, it could either mean the lecture is too simple and they're bored, or it's too fast and complex that students don't understand. I need to interact with the students immediately and ask if they can follow.}''

\par Other methods, such as assignments, exams, and teaching evaluations, serve as post-hoc tools for gathering student feedback, but these often fail to provide timely and specific insights. For example, I2 mentioned, ``\textit{Not every class ends with homework... and the homework doesn't cover everything.}'' I1 added, ``\textit{If homework is done incorrectly, the worst-case scenario is that nothing was learned, but it might as well be due to not reviewing notes in time, it depends.}'' Similarly, I3 noted, ``\textit{After class, even after an hour, students' recollections of their own questions become very vague.}''

\subsubsection{[\textbf{Finding 2}] \textbf{Learning challenges affect the willingness to communicate with instructors}}
\par All instructors interviewed unanimously observed that students with lower academic performance are less likely to initiate communication with them. Similarly, survey data shows a strong correlation between the frequency of difficulties encountered in course learning and the students' willingness to communicate these issues to instructors($r=0.96$, $p<0.01$\footnote{$r$ is the Pearson Correlation Coefficient. We excluded 41 responses from the analysis where participants reported `Never' have comprehension problem and had a `Neutral' stance on their willingness to provide feedback, resulting in a sample size of $n=90$. Also, to improve the sample size, survey responses were categorized into two groups based on the willingness to provide feedback: those willing to provide feedback(`Strongly Disinclined' and `Disinclined') and those unwilling(`Strongly Inclined' and `Inclined').}). Regardless of their inclination to provide feedback, `Feedback mechanism deficiency' (Willing: $37/88$ ($42.0\%$); Unwilling: $24/31$ ($77.4\%$)) and `Self-diagnosis difficulty' (Willing: $42/88$ ($47.7\%$); Unwilling: $11/31$ ($35.5\%$)) were identified as the two primary challenges faced by students.

\par Open-ended survey responses suggest that students prefer having off-public or indirect channels to provide feedback to their instructors ($8/129$, $6.2\%$). This preference aligns with the instructors' observation from the interviews, where they noted that students may hesitate to ask questions in class or directly communicate with instructors due to apprehension or shyness. While instructors often infer students' struggles from their expressions, as I6 noted, ``\textit{Without targeted questions, it is difficult for me to guess where the real problem lies. I either repeat the key points or re-explain based on my understanding... If students want to learn, they need to actively communicate with me. I have tried to probe once or twice, but if there is no response, I believe I have fulfilled my duty.}''

\subsubsection{[\textbf{Finding 3}] \textbf{Expertise in recognizing student understanding}}
\par In interviews, experienced instructors (I1, I2, I4) reflected on how their decades of teaching have built their confidence in identifying common student errors and comprehension difficulties. When faced with unexpected questions, they adeptly use progressive questioning to guide students in uncovering the root of their misunderstandings. As I2 noted, ``\textit{It's not possible to fully grasp what the student is thinking right away; sometimes I really don't understand their questions, but I'll break down the issue into smaller, simpler concepts for confirmation.}''

\par In contrast, novice instructors (I3, I5) expressed more uncertainty regarding student performance and shared feelings of pessimism and helplessness when students encounter learning obstacles. I3 stated, ``\textit{Their backgrounds are so diverse, and they're hesitant to communicate proactively, it's always challenging to gauge the depth and pace of my lectures.}'' I5 mentioned, ``\textit{If students don't understand, I'll explain it again. But if they still don't get it, I'm at a loss for what to do next.}'' Unlike more experienced counterparts, novice instructors tend to place greater emphasis on students' self-study habits and show less empathy in connecting with students.

\subsubsection{[\textbf{Finding 4}] \textbf{Ensuring majority comprehension within teaching constraints}}
\par Instructors work within the constraints of a fixed syllabus, allowing some flexibility to adjust their teaching styles but requiring all content to be covered by the end of the semester. The more detailed the explanation and the more interaction with students, the more time-consuming the process becomes. When faced with a heavy teaching load or tight schedule, instructors often prioritize ensuring the learning experience of students with average and above-average performance. Students with weaker foundational knowledge and understanding are typically categorized as a special group, whose needs are not addressed within the regular teaching plan. As I6 remarked, ``\textit{I don't have the time and energy to delve into their difficulties}''. I5 added, ``\textit{I will announce the basic knowledge used in the course in advance, and students need to fill in the gaps in their spare time.}''

\par Additionally, I3, I4, I5, and I6 emphasized the need for aggregated feedback to better focus on common issues and adjust the teaching content and pace accordingly. I1, I2, I3, and I6 expressed a preference for real-name feedback. When asked for the reason, it was found that, besides high school instructors (I1, I2) needing to track each student's learning progress, instructors generally need to assess how to address problems based on students' background information. For instance, I1 pointed out, ``\textit{Students at different levels have different depths of problems and require different measures.}'' I2 also noted, ``\textit{If a good student makes a mistake, it means most students do not understand my explanation, and I need to adjust.}''

\subsubsection{[\textbf{Finding 5}] \textbf{The impact of prerequisite knowledge on communication}}
\par Survey responses indicate that $80\%$ of students struggle with learning new information due to the influence of prior knowledge. This challenge arises from unfamiliarity with related field ($65/129$, $50.4\%$), gaps in prerequisite courses ($44/129$, $34.1\%$), or forgetting essential basic knowledge ($42/129$, $32.6\%$), making it difficult for them to grasp new concepts. I2 to I6 acknowledged this issue. I2 noted, ``\textit{It greatly affects classroom efficiency and learning outcomes. If students haven't properly grasped the basics, they'll struggle to keep up with what I'm teaching. I’m also seeking methods to address this issue.}''

\par The lack of transparency regarding gaps in prior knowledge between instructors and students, combined with communication barriers, can create significant teaching challenges. I5 shared an example, ``\textit{Once I directly used multivariate Gaussian distribution in my lecture, assuming students to be familiar with it from their stats class, however, students couldn't follow. Later I learned that this distribution had only been briefly introduced, not taught in detail.}''

\par Moreover, when students lack prerequisite knowledge, they often struggle to clearly articulate their difficulties to instructors. I4 observed, ''\textit{It hinders the formation of their knowledge network. They might see there's a problem but can't pinpoint the cause.}'' Students frequently struggle to identify their own knowledge gaps (I2, I3, I4) and often present disorganized questions (I5).

\subsubsection{[\textbf{Finding 6}] \textbf{Embracing online platforms for enhanced learning}}
\par Although instructors acknowledge that online teaching may hinder their ability to observe students' learning status, they also emphasize its benefits, including abundant teaching resources, flexible scheduling and location, a variety of feedback channels, and support for personalized learning. Instructors often integrate features of online education platforms into their offline teaching, including sharing supplementary materials, posting tests, and collecting feedback. However, to use these platforms effectively, instructors must manually configure many functions in advance. Some platforms and tools even require specialized smart classrooms, which can be cumbersome and complex, with high hardware demands, hindering the deep integration of promising TEL tools.

\par Survey results indicate that students are generally willing to use online platforms proactively to mark and communicate content they don't understand (non-anonymous: $93.0\%$, anonymous: $99.2\%$), share their interactions with course videos with instructors (non-anonymous: $82.9\%$, anonymous: $98.4\%$), and utilize TEL tools to facilitate communication with their instructors ($97.7\%$). Offering diverse feedback channels and maintaining anonymity might encourage more interaction between students and instructors.

\subsection{Design Requirements}
\par Based on the six key findings, our work integrates AI methods and visualization strategies into online education platform interfaces tailored for students and instructors. This integration aims to enhance learning environment and feedback loop, providing better support for comprehensive and nuanced analysis of that feedback. The specific design requirements for the student [\texttt{DS}] and instructor end [\texttt{DI}] are outlined below:

\subsubsection{Student End}
\begin{enumerate}
    \renewcommand{\labelenumi}{[\texttt{DS\theenumi}]}
    \item \textbf{Support Multiple Feedback Channels.} According to [Finding 6], Online learning platforms excel at gathering varied student feedback. They enable students to actively comment and question while also implicitly track behavioral patterns. Anonymity in feedback can alleviate students' psychological burden, encourage more proactive responses, and help instructors promptly grasp students' learning status. Additionally, [Finding 1] indicates the student interface should motivate students to provide more detailed feedback.
    \item \textbf{Facilitate Incremental Learning.} Students who struggle with basic concepts often face challenges with advanced material, which hinders their overall subject understanding. According to [Finding 5], the student interface should identify and recommend the prerequisite knowledge needed for each learning activity to support gradual and effective learning progression.
    \item \textbf{Assist Students in Self-Diagnosing Their Knowledge Gaps.} Students lacking prerequisite knowledge or encountering complex explanations may struggle to learn. [Finding 2, 4\&5] show that enabling students to identify the root causes of these challenges helps them resolve issues independently and provide clearer, more precise feedback to instructors.
\end{enumerate}

\subsubsection{Instructor End}
\begin{enumerate}
\renewcommand{\labelenumi}{[\texttt{DI\theenumi}]}
\item \textbf{Automatically Summarize and Organize Student Feedback.} Considering [Finding 4], the system design should ease the burden on instructors by streamlining the collection and analysis of student feedback. It should extract common themes and highlight recurring issues to prevent information overload, taking advantage of the online platform mentioned in [Finding 6].
\item \textbf{Correlate Student Feedback with Lecture Content.} Given that feedback can be delayed [Finding 1], the system should provide relevant contextual information to facilitate precise analysis. Referring to [Finding 3], it should also help narrow down issues to avoid difficulties in tracing problem origins due to factors like poor memory [Finding 5].
\item \textbf{Enhance Teaching Skills Through Retrospective Analysis.} Responding to [Finding 2\&3], the system should support instructors, particularly less experienced ones, in uncovering their unconscious misunderstandings about students and developing empathy towards students.
\end{enumerate}

\section{System}
\begin{figure*}[h]
  \centering
  \includegraphics[width=\linewidth]{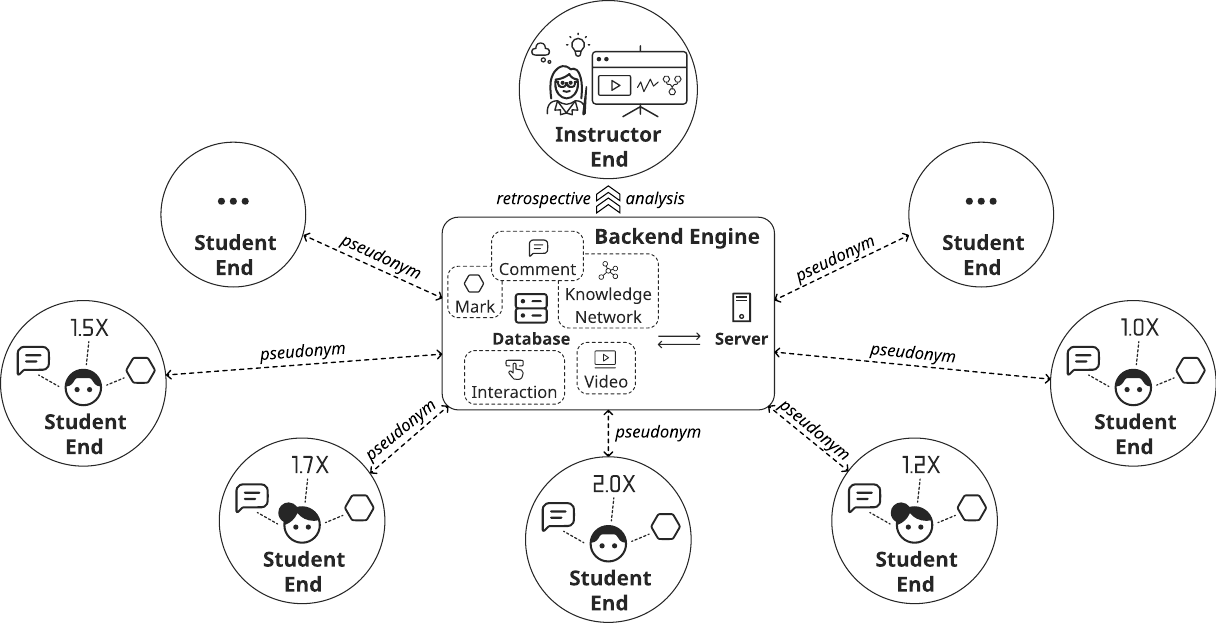}
  \caption{The system architecture includes a central backend engine and dual frontend interfaces: a student end for pseudonym video viewing and feedback, and a teacher end for retrospective analysis insights.}
  \Description{The system architecture.}
  \label{fig:System}
\end{figure*}

\subsection{System Overview and Architecture}
\par
\begin{figure*}[h]
  \centering
  \includegraphics[width=\linewidth]{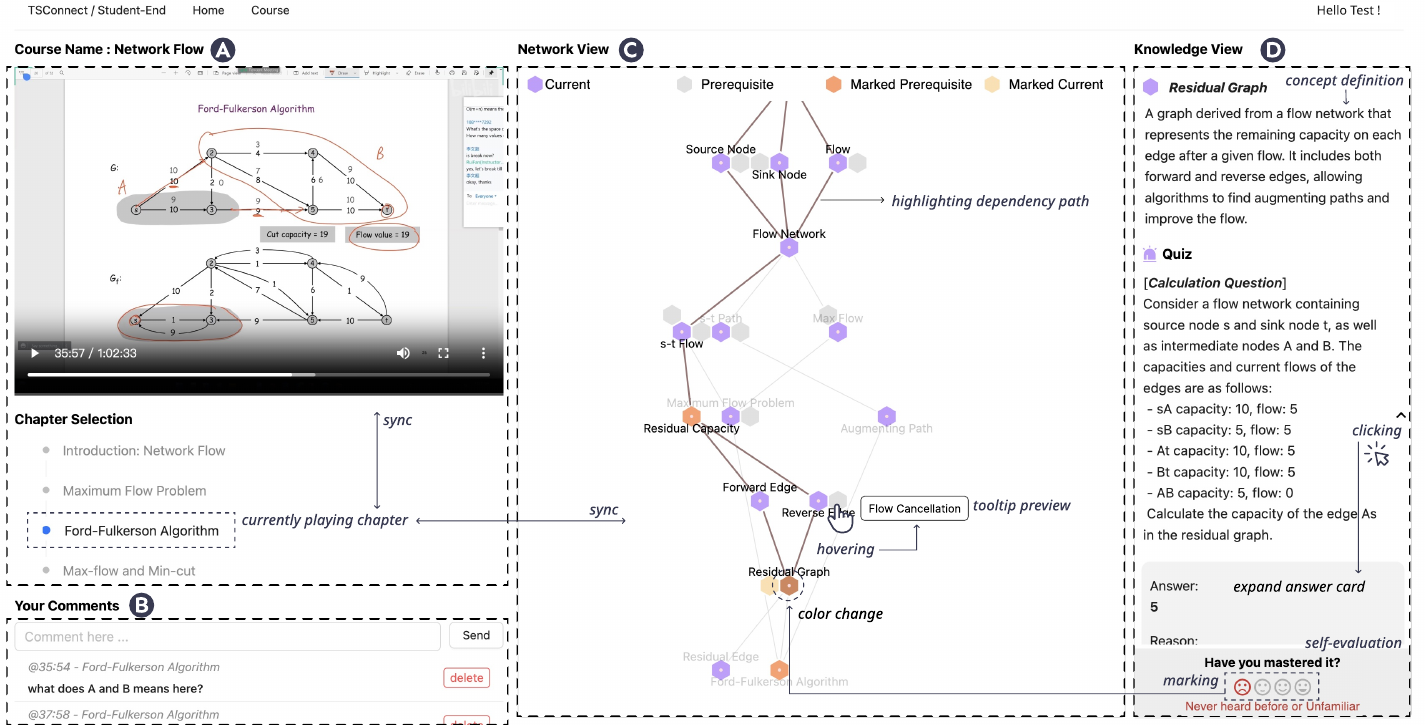}
  \caption{Student end interface of \textit{TSConnect}, featuring: A) the Course Video Player, B) the Comment Section, C) the Network View for displaying prerequisite dependency relationships, and D) Knowledge View for self-evaluation.}
  \label{fig:studentEnd}
\end{figure*}

\begin{figure*}[h]
  \centering
  \includegraphics[width=\linewidth]{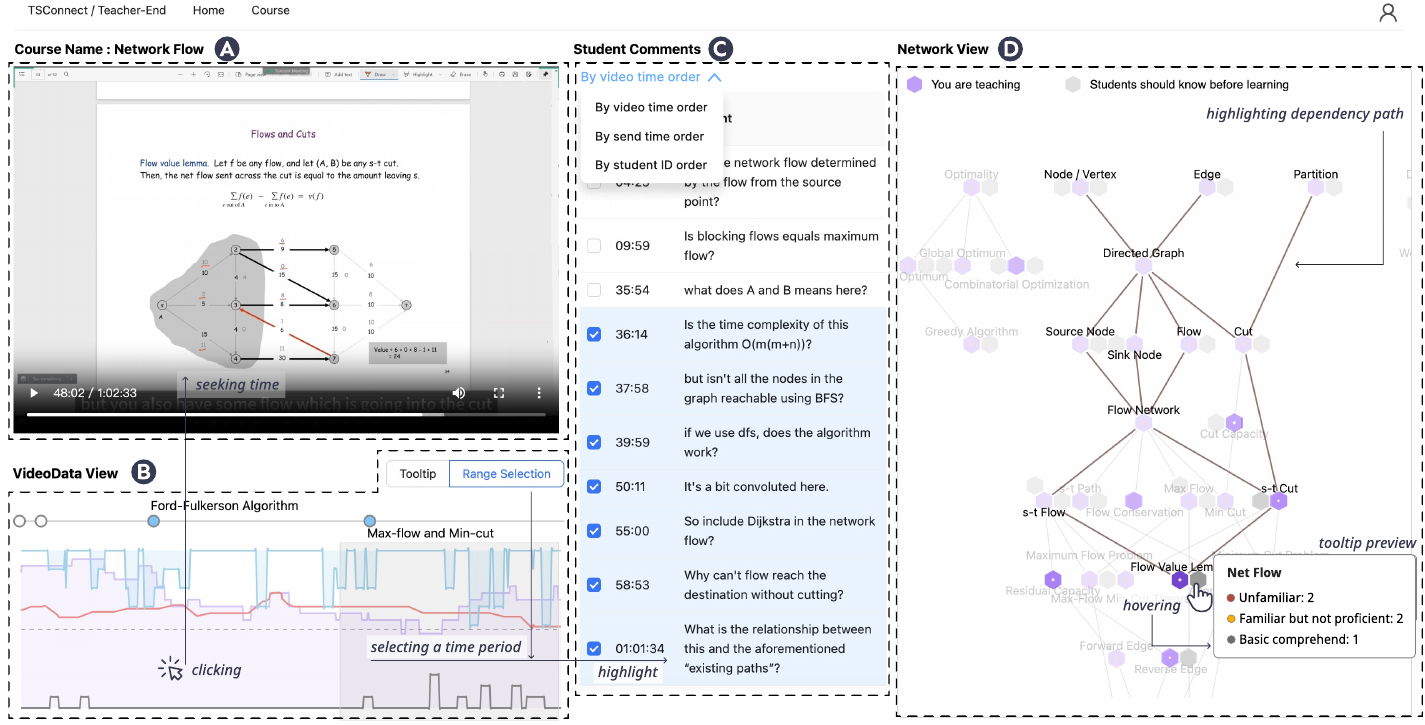}
  \caption{Instructor end interface of \textit{TSConnect}, featuring: A) the Course Video Player, B) the VideoData View, C) the Comment Section, and D) the Network View for displaying prerequisite dependency relationships.}
  \label{fig:instructorEnd}
\end{figure*}

 In line with design requirements [\texttt{DS}]s and [\texttt{DI}]s from our survey and interviews, we proposed \textit{TSConnect}, an interactive online learning system to enhance communication between instructors and students, accessible via PC or tablet. \textit{TSConnect} comprises three main components (\autoref{fig:System}): a backend Engine, a React web-based student end and an instructor end: 1) The back-end engine processes course videos on a Flask server, extracting a knowledge dependency network to establish a feedback channel. All feedback is stored in an SQLite3 database and managed by an Express server. 2) The student end (\autoref{fig:studentEnd}) captures diverse student feedback using pseudonyms for login, uploading the data to the database. 3) The instructor end (\autoref{fig:instructorEnd}) retrieves and visualizes aggregated student feedback, aiding in teaching outcome analysis. The system focuses on enhancing existing feedback mechanisms to improve student engagement and teaching quality, rather than creating a new online education platform. \textit{TSConnect} is designed for seamless integration into any existing online education platform.

\subsection{Video Processing and Graph Construction} \label{sec:VideoProcessing}
\par Upon uploading pre-recorded course videos to the database, instructors can manually annotate chapters. The backend server then processes these annotated videos through the following steps, ultimately generating a knowledge network for students to use on the \textit{TSConnect} learning platform.

\par \textbf{Video Keyframe Extraction}. To alleviate the burden of manually providing written course materials, the server employs an algorithm based on maximum inter-frame difference to automatically detect and extract keyframes from video. These keyframes replace lecture notes and form the basis for identifying and extracting knowledge concepts. The server computes the frame difference between consecutive frames to determine the average pixel-wise difference intensity. Frames with local maxima in this intensity are identified as keyframes. To avoid redundancy, the server smooths the average intensity sequence using a Hanning Window, retaining only one frame from each group of similar keyframes (threshold = $0.9$). The server then employs the PaddleOCR PP-OCRv3\footnote{https://github.com/PaddlePaddle/PaddleOCR} model to perform OCR recognition on each keyframe, generating a text sequence for comparison.

\par \textbf{Knowledge Concept Identification}. Instructors have the option to manually mark multiple chapters within a video upon upload, facilitating the grouping of keyframes. The server processes these keyframes by analyzing the text data chapter by chapter through the GPT-4 Turbo API\footnote{https://platform.openai.com/docs} (temperature=0.4). To enhance the contextual awareness of the language model (LLM) and improve the accuracy of concept extraction, we first require the LLM to identify subtopics within each chapter, followed by the extraction of concepts (termed `course nodes') with prerequisite dependencies closely related to the chapter's topic, rather than conducting frame-by-frame extraction. All course nodes and their relationships from each chapter are unified to create a global set for the entire video, resulting in a comprehensive knowledge dependency graph. In addition to directly merging identical concepts, the server utilizes the Wikipedia API\footnote{https://github.com/goldsmith/Wikipedia} to assist the LLM in resolving concept ambiguities. Furthermore, the server retrieves introductory content from Wikipedia, which is subsequently simplified and refined by the LLM to serve as foundational explanations for the related concepts. Not all extracted knowledge concepts exhibit prerequisite dependencies; for instance, while both `Newton's Second Law' and `Law of Conservation of Energy' rely on `foundational principles of classical mechanics', they are considered parallel knowledge within the dependency graph without direct connections. To prevent isolated nodes after the global set operation, the server instructs the LLM to associate at least one prerequisite concept (referred to as `association nodes') with any course node that has a degree of zero, based on the chapter's theme. For acquiring prerequisite knowledge for each course node, we adopt a straightforward approach: the necessary prerequisite knowledge for each concept should be closely tied to its definition, thus influencing the student's understanding. Consequently, the server extracts hidden prerequisite knowledge from the aforementioned knowledge explanations. If a prerequisite concept has already appeared as a course node or association node, the corresponding course node will be labeled instead of being repeated as an additional prerequisite node.

\begin{figure*}[h]
  \centering
  \includegraphics[width=\linewidth]{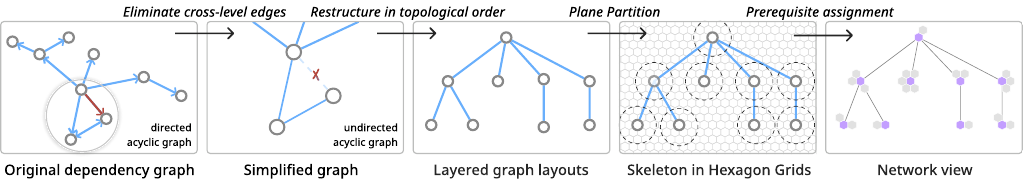}
  \caption{The backend pipeline for dependency graph construction}
  \Description{Figure about Dependency Graph Construction.}
  \label{fig:denseDAG}
\end{figure*}

\par \textbf{Dependency Graph Construction}. The skeleton of the knowledge dependency graph is composed of disambiguated course nodes and association nodes, with directed edges representing the prerequisite relationships between them. We define $G=(V,E)$ as a directed acyclic graph (DAG), where $V$ is a non-empty set of nodes formed by the disambiguated concepts, and $E$ is the set of directed edges representing dependencies between these nodes. For any edge $e\in E$, it connects a pair of nodes $(u,v)$ such that $u$ is a prerequisite for $v$, depicted as $u\rightarrow v$ when understanding or applying $v$ requires prior comprehension of $u$. However, as shown in \autoref{fig:denseDAG}, the initial DAG can be complex and confusing, making it difficult for users to quickly identify prerequisite relationships. To address this issue, the server leverages the transitivity of dependency relations to eliminate redundant cross-level edges that could create cycle structures. Additionally, inspired by the work of \cite{Xie:2021:Visual}, we implement layered graph layouts in topological order and arrange nodes by out-degree from left to right within each layer to minimize edge crossings. Once the skeleton is established, the server employs a hexagonal encoding for all nodes, determines the coordinates for the skeletal nodes, and fills the surrounding space with prerequisite nodes. Given that the average number of prerequisites per skeleton concept is less than $15$, a two-layer hexagonal structure surrounding each skeleton node can accommodate up to 18 nodes. Therefore, we set a minimum distance between skeletal nodes equal to five hexagon side lengths. The server first generates a hexagonal lattice to define the central coordinates of the skeleton nodes, then draws Voronoi diagrams to appropriately fill in the prerequisite knowledge. The resulting knowledge dependency graph will be detailed in \autoref{sec:stuEnd} and \autoref{sec:InstEnd}, which will include specific visualization encoding and interaction mechanisms.

\subsection{Student End} \label{sec:stuEnd}

\subsubsection{\textbf{Course Video Player}} 
\par Building on \cite{Shi:2015:VisMOOC}, we generate second-by-second counts for  play, pause, and rate change events to collect click-stream data. This method effectively communicates students' natural learning behaviors to instructors, acting as a passive feedback channel [\texttt{DS1}] that provides objective contextual information. Similar to conventional MOOC platforms, we include a chapter progress bar beneath the video player to facilitate quick navigation.

\subsubsection{\textbf{Comment Section}} 
\par Students can ask questions, share opinions, and delete previous comments through the \textit{Comments Section} [DS1]. Comments are displayed chronologically with video timestamps and include chapter titles and content.

\subsubsection{\textbf{Network View}} 
\par To support structured learning ~[\texttt{DS2}], we design a \textit{Network view} that visualizes a knowledge dependency subgraph created by the back-end server, as described in~\autoref{sec:VideoProcessing}. This subgraph aligns with the currently playing chapter by removing all non-essential nodes from the global graph-those irrelevant or not prerequisite to the current chapter's concepts. Each node in the view represents a knowledge concept using a hexagonal glyph, with colors signifying attributes. \purpleNode{Purple hexagons} \raisebox{-.38\height}{\includegraphics[height=3ex]{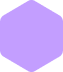}} represent course and association nodes, which form the core structure of the graph and are referenced in the current course video~\footnote{Association nodes are minimally used in the current course video, so they are simplified in the presentation to reduce cognitive load.}. \grayNode{Gray hexagons} \raisebox{-.38\height}{\includegraphics[height=3ex]{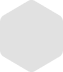}} denote prerequisite nodes, corresponding to concepts not covered in the current video but necessary for understanding the course content. When users interact with knowledge in the \textit{Knowledge View} and mark it, the corresponding \purpleNode{purple} and \grayNode{gray nodes} turn \lorangeNode{light orange} \raisebox{-.38\height}{\includegraphics[height=3ex]{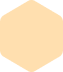}} and \dorangeNode{dark orange} \raisebox{-.38\height}{\includegraphics[height=3ex]{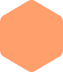}} respectively. Clicking a node highlights the path of dependencies, clarifying knowledge relationships (\autoref{fig:studentEnd}-C). Hovering over a node displays the concept name, while more detailed information appears in the \textit{Knowledge View}.

\par Additionally, when all marked concepts are highlighted in the \textit{Network View}, the resulting topology can serve as an indicator, pinpointing areas where students may be encountering difficulties. This visual representation helps students engage in self-reflection and more effectively summarize their learning challenges~[\texttt{DS3}].



\subsubsection{\textbf{Knowledge View}} 
\par As a complement to the \textit{Network View}, the \textit{Knowledge View} offers more detailed information about individual knowledge concepts, including definitions and corresponding quizzes, which respectively help students reinforce their understanding, and enable self-assessment~[\texttt{DS3}]. Based on student expectations gathered from our formative study (\autoref{App:SelfevaPrefer}), answers and explanations are initially hidden to encourage critical thinking before revealing solutions. At the bottom, a 4-point reflective scoring module allows students to self-evaluate their mastery of the concept~(\autoref{tab:assessLegend}), serving as the third feedback channel in \textit{TSConnect}~[\texttt{DS2}].

\subsection{Instructor End} \label{sec:InstEnd}


\subsubsection{\textbf{Course Video Player}} 
\par
\begin{figure}[ht]
  \centering
  \includegraphics[width=0.55\linewidth]{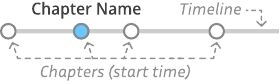}
  \caption{A chapter indicator under the video player.}
  \label{fig:chapterLine}
\end{figure}

 The \textit{Course Video Player} enables instructors to review the original video content~[\texttt{DI3}] with a horizontal chapter progress line (\autoref{fig:chapterLine}) Interactions with the \textit{VideoData View} highlight the current chapter node, linking feedback to the video's sequence~[\texttt{DI2}].

\begin{figure}[h]
    \centering
    \includegraphics[width=\linewidth]{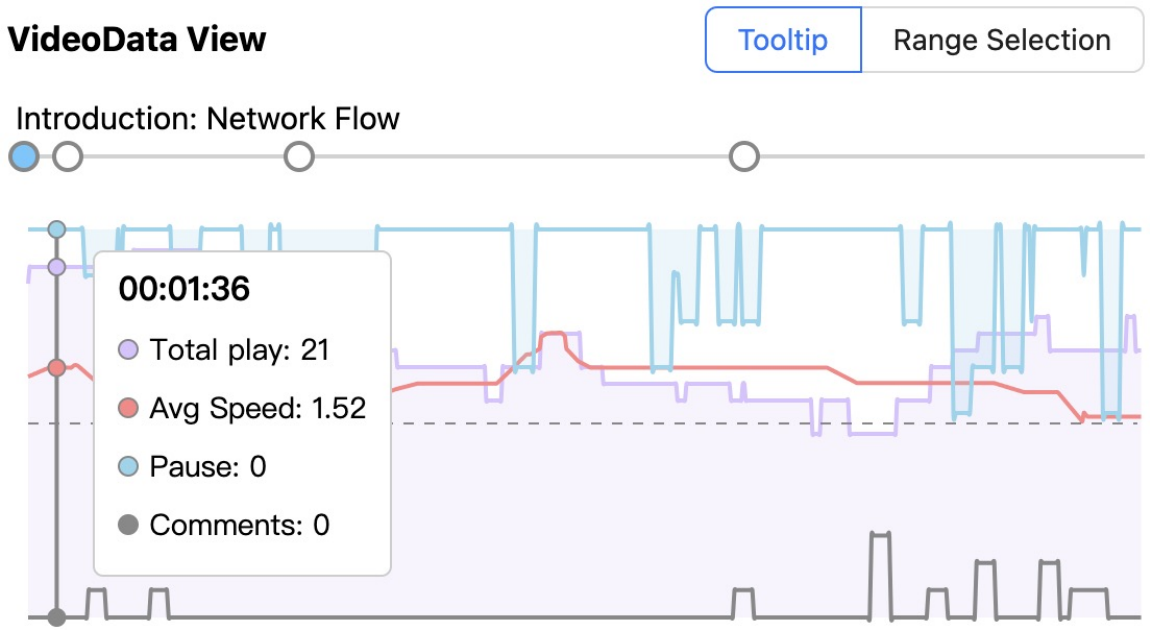}
    \caption{The Tooltip mode of the VideoData View. Upon mouse hover over the view, the system displays detailed feedback statistics while simultaneously highlighting the corresponding chapter title for the given temporal point.}
    \label{fig:feedbackStatistics}
\end{figure}

\subsubsection{\textbf{VideoData View}} 
\par This view organizes key interaction data between students and the course video in chronological order~[\texttt{DI1}], capturing metrics such as total play and pause counts, average playback speed, and the number of comments. Both \raisebox{-.30\height}{\includegraphics[height=3ex]{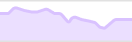}} \purpleNode{plays} (in purple) and \raisebox{-.30\height}{\includegraphics[height=3ex]{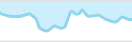}} \pauseArea{pauses} (in blue) are represented as area charts, with plays accumulating from the lower edge and pauses from the upper edge. The \raisebox{-.30\height}{\includegraphics[height=3ex]{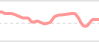}} \speedLine{Speed} (in red) is depicted by a line graph, using the midline as a baseline for $1x$ speed, visualizing playback rate fluctuations across all students. Additionally, \raisebox{-.26\height}{\includegraphics[height=3ex]{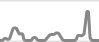}} \grayNode{The number of comments} (in gray) is shown as a line chart growing from the lower edge, representing the cumulative comment count. This intuitive visual representation enables instructors to immediately recognize potential issues in their instruction, guiding them toward targeted exploration and improvements~[\texttt{DI3}].

The \textit{VideoData View} offers two interactive modes: 1) \textit{Tooltip Mode}: Hovering displays detailed feedback statistics for the selected time point~(\autoref{fig:feedbackStatistics}), with the corresponding chapter node highlighted on the chapter timeline. Clicking the node allows the \textit{Course Video Player} to jump to that moment. 2) \textit{Range Selection Mode}: Users can drag to select a time range, which highlights the corresponding chapter node and brings the comments within that range into focus in the \textit{Comment Section}~[\texttt{DI2}].

\subsubsection{\textbf{Comment Section}}
\par Instructors can view student feedback through three sorting options: 1) Sorting by submission time allows instructors to find out the most recent feedback, useful when reusing the same video across multiple student cohorts. 2) Sorting by video timestamp links feedback chronologically to course content, allowing instructors to efficiently locate relevant comments through the \textit{VideoData View} and analyze the feedback in context. 3) Sorting by anonymous student ID enables instructors to track specific issues raised by individual students, facilitating targeted analysis.

\subsubsection{\textbf{Network View}} 
\par The Network View for instructors presents a complete knowledge dependency graph. The backend server calculates overall scores for each concept by aggregating students' self-evaluation scores, ranging from $0$ (Never Heard or Unfamiliar) to $3$ (Completely Mastered)~(\autoref{tab:assessLegend}), then mapped into corresponding color intensity on instructor end. Nodes darken with more feedback, particularly for concepts with commonly weaker mastery. By visualizing the distribution of these scores across the knowledge dependency graph, instructors can easily identify common areas where students face difficulties~[\texttt{DI2}]. Additionally, the relationships between knowledge nodes help instructors analyze potential root causes, fostering more empathy towards students~[\texttt{DI3}]. For example, they may realize whether they have overlooked students' understanding of prerequisite concepts, or whether challenges stem primarily from the current knowledge being taught.

\begin{table}[ht]
  \begin{tabular}{rcl}
  \toprule
  Score                   & Icon                             & Description  \\ \hline
  \color[HTML]{FF4D4F}{3} & \raisebox{-.18\height}{\includegraphics[height=2.3ex]{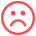}}  & \color[HTML]{FF4D4F}{Never heard before or Unfamiliar} \\
  \color[HTML]{FAAD14}{2} & \raisebox{-.18\height}{\includegraphics[height=2.3ex]{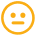}}  & \color[HTML]{FAAD14}{Familiar but not Proficient} \\
  \color[HTML]{8C8C8C}{1} & \raisebox{-.18\height}{\includegraphics[height=2.3ex]{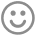}}  & \color[HTML]{8C8C8C}{Basic Comprehend} \\
  \color[HTML]{52C41A}{0} & \raisebox{-.18\height}{\includegraphics[height=2.3ex]{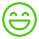}}  & \color[HTML]{52C41A}{Completely Mastered}     \\ \bottomrule
  \end{tabular}
  \caption{A legend and conversion rule for the scoring module in the \textit{Knowledge View} in Student end.}
  \label{tab:assessLegend}
\end{table}


\begin{table*}[ht]
    \centering
    \begin{tabular}{ccccc|ccccc}
    \toprule
    \multicolumn{5}{c|}{TSConnect}                               & \multicolumn{5}{c}{Baseline}                                 \\ 
    ID   & Gender & Age & Role          & Major                  & ID   & Gender & Age & Role          & Major                  \\ \hline
    PS1  & Female & 19  & Undergraduate & Computer Science       & PS2  & Male   & 23  & Graduate      & NLP                    \\
    PS3  & Male   & 23  & Graduate      & HCI                    & PS4  & Female & 26  & Graduate      & Data Science           \\
    PS5  & Female & 25  & Graduate      & HCI                    & PS6  & Female & 21  & Undergraduate & Computer Science       \\
    PS7  & Female & 21  & Undergraduate & Computer Science       & PS8  & Male   & 24  & Graduate      & Biomedical Engineering \\
    PS9  & Male   & 25  & Graduate      & NLP                    & PS10 & Male   & 21  & Undergraduate & Computer Science       \\
    PS11 & Male   & 20  & Undergraduate & Computer Science       & PS12 & Female & 22  & Undergraduate & Computer Science       \\
    PS13 & Female & 23  & Graduate      & NLP                    & PS14 & Female & 24  & Graduate      & HCI                    \\
    PS15 & Male   & 22  & Undergraduate & Electronic Engineering & PS16 & Male   & 25  & Graduate      & Data Science           \\
    PS17 & Male   & 21  & Undergraduate & Computer Science       & PS18 & Male   & 20  & Undergraduate & Electronic Engineering \\
    PS19 & Male   & 24  & Graduate      & Computer Vision        & PS20 & Female & 25  & Graduate      & NLP                    \\
    PS21 & Female & 26  & Graduate      & Biomedical Engineering & PS22 & Male   & 22  & Undergraduate & Computer Science       \\
    PS23 & Male   & 22  & Undergraduate & Biomedical Engineering & PS24 & Female & 20  & Undergraduate & Computer Science       \\
    PS25 & Male   & 26  & Graduate      & Robotics               & PS26 & Male   & 27  & Graduate      & HCI                    \\
    PS27 & Female & 23  & Undergraduate & Computer Science       & PS28 & Male   & 22  & Undergraduate & Computer Science       \\
    PS29 & Female & 24  & Graduate      & Data Science           & PS30 & Female & 23  & Graduate      & Mathematics            \\ \bottomrule
    \end{tabular}
    \caption{\reviewAdd{Student information in user study.}}
    \label{tab:userStudyDemo}
    \end{table*}

\section{User Study}
\par To address \textbf{RQ3} and \textbf{RQ4-a}, we conducted a between-subjects user study with 30 student, following institutional IRB approval. Students participated in one professional course session using the proposed \textit{TSConnect} system, with a baseline system as control. Additionally, we interviewed 4 course-related instructors, using the feedback data from \textit{TSConnect}, to explore \textbf{RQ4-b} and \textbf{RQ5}. 

\subsection{Conditions}
\par We performed a comparative analysis between the student interface of \textit{TSConnect} and a baseline system, which represents a traditional MOOC platform with basic features like video lecture playback and a text-based comment section. Unlike \textit{TSConnect}, the baseline system lacks two key components: the \textit{Network View} and the \textit{Knowledge View}. Additionally, participants using the baseline system were provided unrestricted access to external knowledge sources, such as Wikipedia and other online encyclopedias.

\subsection{Participants}
\par Following approval from the university's IRB, we recruited 30 students enrolled in an algorithm analysis course at a local university. The participants, comprising 16 male and 14 female students with an average age of 22.9 (SD = 4.1), included 14 senior undergraduates and 16 graduate students. Participants were randomly assigned to either the baseline system or \textit{TSConnect}, based on demographic factors and their learning preferences\footnote{Learning preferences include students' academic proficiency, their inclination to seek instructor guidance when facing learning challenges, and their tendency for autonomous learning.}. The experimental materials consisted of video lectures recorded during the COVID-19 pandemic, covering topics from the latter half of the course curriculum. Recruitment occurred early in the academic semester, and we verified that none of the participants had prior exposure to these materials, ensuring that the experimental content was independent of the material covered in the first half of the course. Upon completion of the student experiments, we populated the instructor interface of \textit{TSConnect} with all collected feedback data. We then conducted semi-structured interviews with four faculty members (PI$1\sim4$, three males and one female, average age of 35.4) who teach the algorithm course at the local university. Together with the instructors, we explored the instructor interface of \textit{TSConnect}. The entire study lasted approximately one hour for student participants and 30 minutes for instructor participants. Instructors and students were compensated USD $8$ and USD $5$, respectively.

\subsection{Task and Procedure}
\subsubsection{Task} 
In this study, participants were assigned to use either the baseline system or \textit{TSConnect} to engage with the same video lecture on Network Flow. Participants were granted full control over video playback, including variable speed settings replay and skip. However, they were instructed to maintain focus throughout the session, refraining from external communication or engagement in unrelated activities. To incentivize engagement, participants were informed that their compensation would be contingent upon their performance in a post-study quiz (not actually exist). We encouraged, but did not mandate, the use of the system's feedback mechanisms for communicating with instructors. Participants were assured this wouldn't affect their compensation, but we emphasized that their input would help improve future course versions.

\subsubsection{Procedure} 
Before the study, student participants signed a consent form and completed a pre-task demographic questionnaire. We introduced the experimental task and system usage for each condition. To gather more data, both groups were demanded to mark all \purpleNode{skeleton knowledge} in the last chapter. Students using \textit{TSConnect} used the scoring module in the \textit{Knowledge View}, while those using the baseline system completed a self-assessment form with the same criteria. Subsequently, all student participants completed a post-task questionnaire. Two of the authors acted as experimenters to ensure smooth progress and provided assistance as needed.

\subsection{Measurement}
\par 
We designed a 7-point Likert scale (1: Not at all/Strongly disagree, 7: Very much/Strongly agree, and a 10-point scale for workload-related questions) post-task questionnaire to collect student participants' experience on the respective systems. First, we crafted questions on \textbf{Usability} of the system referring the System Usability Scale (SUS) including 1) Ease of use; 2) Learning support; 3) System satisfaction; 4) Likelihood of future use. Second, referring to the NASA-TLX survey~\cite{Hart:1988:Development}, we propose questions for the effects on students' \textbf{workload} including 1) Cognitive load; 2) Workload; 3) Frustration level; 4)Performance. Third, in terms of \textbf{Learning Behavior}, we design questions including 1)Encountered learning difficulties; 2) Feedback willingness; 3)Clear problem identification; 4) Problem resolution; 5) More feedback than usual. Fourth, as for \textbf{System Design}, we tailored questions concerning the \textit{Network View} and \textit{Knowledge View} for participants using \textit{TSConnect}, including: 1) Intuitive visualization; 2) Convenience of interaction; 3) Overall helpfulness; 4) Mechanism Approval. Additionally, we also included optional subjective questions for qualitative insights. While the instructor end utilized final scores for retrospective visual representation, the system backend server logged each score modification made by student participants. These granular operational data provided crucial support for subsequent analyses.

\begin{figure*}[ht]
  \centering
  \includegraphics[width=\linewidth]{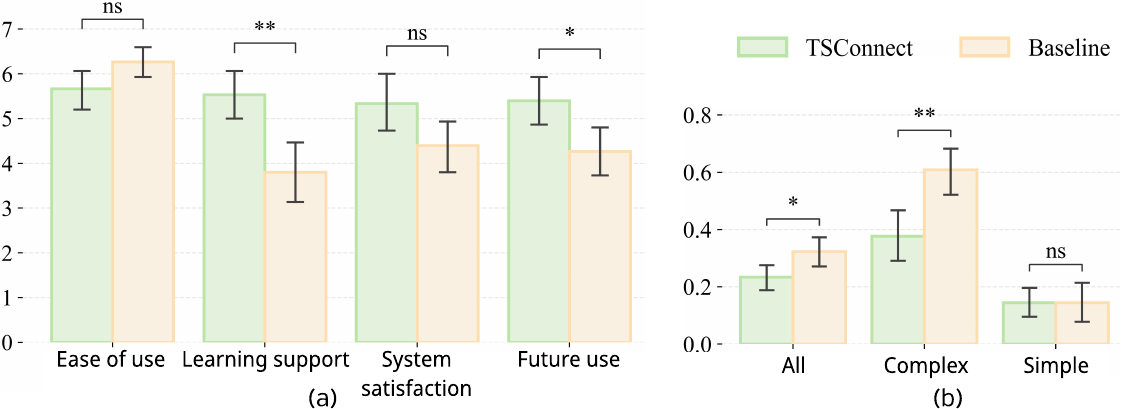}
  \caption{Results of the (a) usability of usefulness of the system and (b) differences in self-evaluation score results among participants after using different systems. The error bars indicate standard errors. (ns: p < .1; $^{\ast}$: p < .05; $^{\ast\ast}$: p < .01)}
  \label{fig:RQ3}
\end{figure*}


\section{Results and Analysis}
\par 
For quantitative analysis, we employed the Mann-Whitney U test~\cite{Mann:1947:Test} on post-task questionnaires responses besides descriptive statistics. For qualitative analysis, instructors reviewed the student feedback by \textit{TSConnect} in the interview. We explored instructors' perception of feedback data in each system view and implications for their future teaching. Two researchers independently coded interview transcripts, followed iterative discussions to reach consensus for thematic analysis~\cite{Guest:2011:ThematicAnalysis}.

\subsection{RQ3: What is the usability and effectiveness of the support system?}\label{sec:RQ3}

\par
As shown in~\autoref{fig:RQ3}-(a), the survey results presents participant ratings of system usability with different systems. Our analysis indicates that \textit{TSConnect} did not result in statistically significant changes in `Ease of Use' or `System Satisfaction'. However, it did demonstrate a significant enhancements in `Learning Support' (U = 188, p < 0.01) and `Future Use' (U = 175, P < 0.05). To evaluate the efficacy of \textit{TSConnect} in facilitating learning, we conducted an analysis of the collected mark data. This analysis uncovered the following two primary findings.

\subsubsection{[\textbf{Finding 7}] \textbf{The \textit{Network View} and \textit{Knowledge View}, significantly enhanced students' capacity to overcome learning obstacles.}} We analyzed the knowledge marking logs from participants using \textit{TSConnect}, the results revealed instances of score modifications with extended time intervals (exceeding $10$ seconds), with a trend towards lower scores after these modifications (occurrences per participant: M = $0.91$ , SD = $0.78$). This phenomenon may indicate that participants gradually deepened their understanding of the relevant knowledge while using the system. To isolate the potential effects of course progression itself, thereby more accurately evaluating the unique contribution of the \textit{TSConnect} system, we further comparatively checked the knowledge self-assessment data from both participant groups.

\par
After the experimental tasks, both participant groups evaluated $26$ \purpleNode{skeleton knowledge} items from the last session chapter. Our analysis goal was to assess how introducing prerequisite relationships and revealing hidden prerequisites affects  learning outcomes. We categorized knowledge based on their prerequisite relationship complexity, which was determined by the sum of quantities of incoming edges in the knowledge network (representing explicit prerequisites), and \grayNode{hidden prerequisites}. The top $40\%$ were classified as `complex', with the rest as `simple'. Subsequently, we calculated the average scores for participants from both groups across these two categories. As illustrated in \autoref{fig:RQ3}-(b), participants using \textit{TSConnect} demonstrated superior overall knowledge mastery (U = $64$, p < $0.05$) compared to the baseline condition (reflected in lower scores) especially for `complex' knowledge (U = $40$, p < $0.01$). These finding suggests that the prerequisite assistance provided by \textit{TSConnect} effectively helped students elucidate the interconnections between knowledge concepts, enabling them to systematically deconstruct and comprehend complex concepts, thereby fostering a more structured learning process.

\begin{figure*}[ht]
  \centering
  \includegraphics[width=\linewidth]{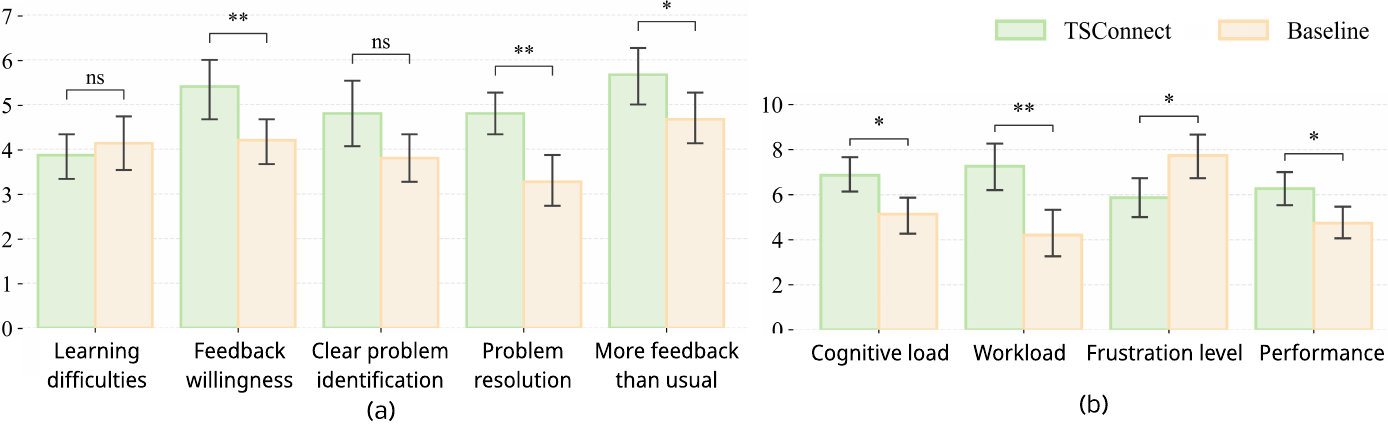}
  \caption{Results of (a) the effect of different systems on learning behavior, and (b) the effect on students' cognitive load, workload, students' perceived level of task-related frustration, and the self-evaluation of their learning performance. The error bars indicate standard errors. (ns: p < .1; $^{\ast}$: p < .05; $^{\ast\ast}$: p < .01)}
  \label{fig:RQ4}
\end{figure*}

\subsubsection{[\textbf{Finding 8}] \textbf{\textit{TSConnect} effectively enhances student-teacher interaction, significantly increasing the amount of proactive feedback from students.}} We conducted a quantitative analysis of feedback data from both groups. Results indicate that the baseline group provided slightly more text-based feedback through the \textit{Comment Section} (M = $1.87$) compared to the \textit{TSConnect} group (M = $1.53$), though this difference was not statistically significant (p > 0.05). Furthermore, participants using \textit{TSConnect} marked an average of $2.53$ knowledge (SD = $1.64$).
\par 
The \textit{Network View} and \textit{Knowledge View} in \textit{TSConnect} collectively constituted an additional feedback channel. However, these new channel did not significantly reduce the utilization of existing text-based feedback. This may be attributed to the fact that text-based feedback can encompass a broader range of complex information, such as evaluations of instructor explanations, which cannot be fully captured by a simple marking mechanism. Concurrently, the operational simplicity of the marking mechanism (requiring only a click to indicate comprehension level) proved more efficient than composing text-based feedback, thereby implicitly lowering the obstacle for student-teacher communication. Questionnaire results indicate that on a $7$-point Likert scale, participants found the design of \textit{Network View} and \textit{Knowledge View} to be intuitive (M = $5.37$, SD = $1.51$), with simple and user-friendly interactions (M = $5.73$, SD = $0.92$). Notably, all participants expressed support for the use of the marking mechanism for feedback (M = $5.48$, SD = $1.04$). 
An in-depth analysis of students' perspectives on these diverse feedback channels will be presented in \autoref{sec:RQ4a}.

\subsection{RQ4-a: How do students perceive the support system?}\label{sec:RQ4a}
\par We conducted a comprehensive analysis of both quantitative scales and open-ended questions from the questionnaire, aiming to thoroughly investigate the impact of \textit{TSConnect} on student participants' workload and their learning performance.

\subsubsection{Effects on students’ workload.} \autoref{fig:RQ4}-(b) visually compares the workload differences between the baseline and \textit{TSConnect} group. Results reveals that \textit{TSConnect} significantly increased both the cognitive workload (U = $171$, p < $0.05$) and overall workload (U = $189$, p < $0.01$) for students completing learning tasks. This increase could be attributed to the rich features and content provided by \textit{TSConnect}, which required participants to interact extensively with the system, engaging with both textual and graphical information beyond just watching videos.

\par Despite the increased workload, \textit{TSConnect} group reported significantly lower frustration level when completing learning tasks (U=$54$, p<$0.05$). Their self-evaluation of the overall learning performance was superior to that of the baseline group (U=$172$, p<$0.05$). These insights suggest that the [\textbf{Finding 9}] \textbf{increased cognitive engagement may lead to a more positive learning experience and improved self-perceived learning outcomes}.

\subsubsection{Effects on students’ learning performance.}
\autoref{fig:RQ4}-(a) presents a comparative analysis of learning behaviors between \textit{TSConnect} and baseline groups. The data indicates that both groups perceived similar levels of difficulty in completing the learning tasks. However, in terms of feedback behavior, \textit{TSConnect} group demonstrated a notable advantage. Compared to their usual feedback patterns, \textit{TSConnect} group showed an increase in both the quantity (U=$162$, p<$0.05$) and willingness (U=$177$, p<$0.01$) to provide feedback to instructors during this experimental task, significantly surpassing the baseline group. This finding highlights the potential value of \textit{TSConnect} in fostering student-teacher interaction. Although no significant difference was observed between the two groups in the dimension of `helping to clarify personal problem', \textit{TSConnect} group reported an enhanced ability to independently resolve issues during the learning process (U=$185$, p<$0.01$). This result aligns with [Finding 1], further supporting the positive role of \textit{TSConnect} in cultivating students' autonomous learning capabilities.

\subsubsection{Participants' opinion on system design.} 
We conducted a thematic analysis of the TSConnect group's responses to open-ended questions in the post-task questionnaire. The results revealed that:
\begin{itemize}
    \item $7$ out of $15$ participants provided positive evaluations of the prerequisite dependency paths in the \textit{Network View}, including `Intuitiveness'($5$), `Step-by-step Learning'($2$), `Structured Knowledge'($4$) and `Attention Allocation'($1$). 
    \item $4$ out of $15$ participants appreciated the definitions and quizzes in \textit{Knowledge View} as as helpful learning supplements. One student participant noted, ``\textit{Quizzes are an effective learning method. I usually reinforce my understanding through post-class exercises. TSConnect integrates this directly into MOOC learning, making knowledge consolidation more timely.}''.
    \item $2$ out of $15$ participants innovatively utilized the marking mechanism as a learning reminder tool besides the original feedback role. One participant reported marking concepts when encountering difficulties in immediate comprehension during initial MOOC video viewing. Another participant marked concepts that proved challenging during quizzes. These opinion shows that the marking mechanism allows students to prepare for subsequent in-depth understanding without interrupting their current learning flow.
\end{itemize}

\subsection{RQ4-b: How do instructors perceive the support system?}\label{sec:RQ4b}
\par In the interviews, we guided four instructor participants to engage with the instructor end of \textit{TSConnect} and explore student feedback data. This process aimed to evaluate the system's functionality and potential impact from the instructor's perspective. Results of the thematic analysis reveals two following findings.

\subsubsection{[\textbf{Finding 10}] \textbf{\textit{TSConnect} increased the quality and interpretability of student feedback}} All four participating instructors agreed that feedback collected by \textit{TSConnect} was clearer and more comprehensible than traditional methods. Specifically, \textit{TSConnect} enables instructors to analyze feedback within the context of course and playback data, particularly play and pause behaviors, which are indicators of student engagement and effort. The emphasized nodes in the \textit{Network View} intuitively display students' grasp of various knowledge concepts, making it easier for instructors to identify common challenges. PI$3$ said, ``\textit{Previously, I'd simply answer questions without much thought or adjusting future lessons. But with TSConnect, I easily spot repeated issues, prompting me to consider their causes and interconnections.}'' Furthermore, \textit{TSConnect} encourages students to provide more specific and focused feedback. As PI$2$ noted: ``\textit{Students no longer merely request general explanations, but can clearly indicate which particular property or derivation step they need detailed clarification on.}''

\subsubsection{[\textbf{Finding 11}] \textbf{\textit{TSConnect} enhances instructors' ability to diagnose root causes of learning obstacles}} During the interviews,  teachers interacted with \textit{TSConnect} to explore potential factors contributing to students' learning difficulties below surface-level feedback. For example, PI$4$ discovered an increase in replay frequency during the $42\sim44$ minute interval. Upon examination, the instructor found that this segment focused on explaining ``Cut Capacity'' concept. Interestingly, the \textit{Network View} displayed a light-colored node for this knowledge, suggesting a high level of student comprehension. PI$4$ re-evaluated the video segment and identified potential issues with the instruction, especially the unclear mark in the figure. This likely contributed to student confusion at initial. Similarly, PI$2$ identified that the concept of ``Net Flow'' is inadequately explained, which serves as a hidden prerequisite in the \textit{Network View}. This instructional deficiency may hinder students' comprehension of the teaching goal ``Flow Lemma''.

\subsection{RQ5: What impact does the support system have on current teaching and learning practices?}
\par Beyond generating insights specific to the experimental course videos, the interaction with \textit{TSConnect} also provided valuable inspiration for enhancing current pedagogical practices. Moreover, it prompted instructors to critically evaluate their established teaching methodologies. Here are three potential impacts of \textit{TSConnect}.

\subsubsection{\textbf{Impact 1: Avoid making and break strong assumptions about students' prior knowledge.}} Instructor often possess a more extensive knowledge base than their students, which can inadvertently lead to the the use of unfamiliar concepts during instruction. This is the cognitive defect brought about by the curse of knowledge, and is difficult for teachers to identify and solve through their own efforts. As discussed in \autoref{sec:formativeFinding}, in existing teaching process students rarely explicitly express that they have encountered problems. \textit{TSConnect} addresses this issue by fostering student-teacher communication regarding learning challenges, potentially reduces the time required for instructors to realize and identify the knowledge gaps, thereby accelerating the development of pedagogical expertise. Furthermore, it enhances instructors' understanding of their student cohort and cultivates empathy. PI$2$ and PI$4$ highlighted an additional benefit of the \textit{Network View} feature within \textit{TSConnect}. Even without feedback data, this dependency graph provides a valuable framework for instructors to proactively assess the prerequisite knowledge of current learning objectives in advance, helping them identify and address potential gaps that could lead to cascading effects before they appear in the classroom.

\subsubsection{\textbf{Impact 2: Iterate and refine the long-term reusable course materials and explanations.}} The instructors participating in this study are engaged in ongoing instructional responsibilities for established courses. Except the initial offering of a course necessitates overall slide preparation and content planning, subsequent iterations typically involve tiny updates based on prior teaching experiences. This approach is inherently subjective and susceptible to memory biases. \textit{TSConnect} addresses these limitations by facilitating the systematic collection of targeted feedback data. It enables instructors to access and review student responses continuously, supporting targeted data-driven refinements to course materials. Similar to the impact of prerequisite, contextual information also influences student comprehension, as PI$4$ identified issues related to inadequate figure marking in \autoref{sec:RQ4b}. \textit{TSConnect}'s functionality allows for post-session analysis, enabling timely identification and rectification of such issues, thereby mitigating potential confusion for future students. PI$4$ added, ``\textit{It's better to reduce unnecessary cognitive load for students, allowing them to focus on more complex concepts requiring deeper engagement.}'' PI$1$ also mentioned this perspective, ``\textit{Sometimes during lectures, I suddenly come up with a better way to explain something. However, without prior preparation, these last-minute changes can lead to disorganized delivery and missed some key points. I know this can hurt student understanding, but it's hard to spot these issues in the moment, and I often forget to address them afterward. A tool like this would help me improve my teaching methods later on.}''

\subsubsection{\textbf{Impact 3: Adopt a critical and selective approach when utilizing the extensive array of MOOC resources.}} PI$3$, a relatively novice instructor, reported regularly reviewing diverse MOOC videos for pedagogical inspiration. However, PI2 acknowledged the limitations of this approach, ``\textit{The efficacy of instructional methods is actually determined by student reception. Unfortunately, without implementing these techniques in my own classroom, it's challenging to accurately assess their effectiveness.}'' This underscores the potential value of enhancing existing MOOC platforms with advanced analytics tools for instructors. By video engagement metrics and knowledge score visualizations, instructors could better evaluate existing MOOC resources, discerning between effective and worse segments within each video to facilitate a dual-pronged approach: adopt exemplary teaching practices and avoid of common pedagogical pitfalls. Moreover, this data-driven approach would offer instructors a broader perspective on typical student challenges across various MOOCs, leading to more realistic expectations of students and ultimately enhance the student learning experience.

\section{Discussion and Limitation}
\subsection{Generalizability}
\par \textit{TSConnect}'s initialization process can be expanded to incorporate not only video content but also slide presentations. This expansion is feasible due to the fundamental similarity in data processing procedures for both media types. Furthermore, by pre-extracting knowledge dependency graphs from slides and leveraging advanced streaming capture and processing technologies, \textit{TSConnect}'s applicability can extend beyond MOOCs to encompass real-time instructional settings, such as live-streamed lectures. This enhancement significantly broadens the system's potential deployment across diverse educational contexts.
\par In the extraction of prerequisite knowledge, our methodology prioritized definition content over property descriptions of concepts. This approach was adopted in recognition of the varying depths and breadths of conceptual understanding required at different educational levels, such as secondary and tertiary education. Additionally, we deliberately limited our extraction to immediate prerequisites, refraining from multi-level prerequisite relationships. We assume that secondary and deeper prerequisites often fall outside the immediate scope of a given lesson. When students identify gaps in their foundational knowledge, they should seek supplementary courses or materials. Also, instructors are not required to closely track students' mastery of these distant prerequisites.

\subsection{System Design}
\par Conventional learning materials such as textbooks and lecture slides are readily accessible and independent of specific hardware or teaching scenarios. These materials have been widely utilized to identify in students' post-class knowledge mastery gaps by previous work~\cite{Bauman:2018:Recommending,Okubo:2023:Adaptive} and instructors. This approach does not constitute an innovation in our work. Instead, the core innovation of \textit{TSConnect} system lies in leveraging these established resources to bridge teacher-student communication. This enables both parties to align student learning needs during the regular teaching process without introducing additional workflow requirements. Beyond merely facilitating problem resolution, \textit{TSConnect} aims to reveal the root causes of learning difficulties through student feedback data, including necessary prerequisite and teaching process, thereby enhancing instructors' understanding of their students and promoting critical reflection on existing pedagogical practices.

\par Beyond validating the utility of the \textit{TSConnect} through user studies, we garnered valuable insights for future enhancements. A key improvement area is integrating three distinct feedback mechanisms into a more cohesive system. For example, we could enhance the textual feedback feature with natural language processing to automatically identify and tag specific knowledge concepts. These tags could be incorporated into the Network View using a scoring conversion rule, enabling instructors to filter feedback by knowledge concepts for targeted analysis. Furthermore, aligning knowledge node markings with video content by timestamp would help instructors pinpoint recurring concepts and their contextual challenges throughout the course progression. Expanding annotation options for knowledge nodes beyond simple scoring could also provide a deeper understanding of student learning needs.
\par Currently, \textit{TSConnect} restricts students to viewing only their own comments to reduce inhibition from peer feedback. However, expanding user privileges to include broader access and peer discussions may be necessary. To deal with this potential modification while maintaining the integrity of individual feedback, we could implement a weighted comment mechanism that students would have the option to endorse existing comments, increasing their significance within the system. This feature offers an alternative metric for assessing feedback prevalence and impact. On the instructor end, endorsed comments could be highlighted using advanced data visualization techniques, enabling educators to quickly identify high-impact feedback.

\subsection{Limitation}
\par This study has several limitations. First, \textit{TSConnect}'s data processing capabilities encounter challenges when applied to MOOC videos that involve extensive handwritten board work. These difficulties arise from multiple factors: 1) Optical Character Recognition struggles with varied handwriting styles. 2) Perspective distortions of board content due to the camera's positioning. 3) Frequent occlusions caused by instructor movement. A potential solution to address these issues involves incorporating audio processing capabilities. This could begin with Automatic Speech Recognition to transcribe the instructor's speech, followed by Natural Language Processing techniques to extract key knowledge concepts from the transcript. However, this audio-based approach was not implemented or assessed in the current study. Second, the quizzes in the \textit{Knowledge View} are generated autonomously by a LLM, which can sometimes result in misalignment between the quiz focus and the intended conceptual assessment, incorrect answers, or unsolvable questions. Future improvements could refine this feature by integrating Retrieval-Augmented Generation (RAG) methods that utilize established question banks. However, direct indexing of matching questions may not be straightforward. Third, the current implementation of the \textit{Knowledge View} primarily emphasizes concept definitions, neglecting detailed properties of those concepts. In practice, a student’s ability to comprehend and apply a concept’s properties often serves as a more accurate indicator of their learning progress than merely understanding its definition. Future iterations could enhance the system by integrating more comprehensive property-based assessments to better capture students' mastery levels.

\par Additionally, the formative study revealed that while instructors are aware of the curse of knowledge phenomenon, they persistently struggle to overcome this cognitive bias due to challenges in improving teacher-student interactions. In response, we developed \textit{TSConnect} with an emphasis on facilitating student feedback, and subsequent user study validated its effectiveness from the student perspective. However, due to time constraints, we were unable to implement \textit{TSConnect} in real educational settings, limiting our ability to evaluate whether long-term usage could effectively mitigate the curse of knowledge and enhance teachers' pedagogical competencies and empathy towards students. A comprehensive assessment of these aspects would necessitate the integration of additional educational metrics and longitudinal research methodologies.

\section{Conclusion}
\par We present \textit{TSConnect}, an adaptable interactive MOOC learning system designed to bridge the communication gap between students and instructors, reducing the obstacles instructors face in addressing the cognitive bias known as the curse of knowledge in their teaching practices. Our contributions are summarized as follows. First, we conducted an exploratory survey and semi-structured interviews to identify the key factors and practical challenges that hinder current educational practices from mitigating this cognitive bias. Based on these insights, we designed and implemented \textit{TSConnect}, which integrates three feedback channels: playback behavior tracking, textual comments, and knowledge concept marking. The system also visualizes prerequisite relationships between knowledge concepts, uncovering hidden prerequisites that promote more structured learning. Third, we conducted a between-subjects user study with 30 students and interviewed four instructors to evaluate the effectiveness of our design. We explored how both students and instructors perceive the system in a simulated MOOC learning task and examined its potential impact on pedagogical practices. Our findings indicate that \textit{TSConnect} encourages students to provide more frequent and clearer feedback, improving instructors' understanding of student learning progress.

\begin{acks}
\par We thank anonymous reviewers for their valuable feedback. This work is supported by grants from the National Natural Science Foundation of China (No. 62372298), Shanghai Engineering Research Center of Intelligent Vision and Imaging, Shanghai Frontiers Science Center of Human-centered Artificial Intelligence (ShangHAI), and MoE Key Laboratory of Intelligent Perception and Human-Machine Collaboration (KLIP-HuMaCo).
\end{acks}







\bibliographystyle{ACM-Reference-Format}
\bibliography{sample-base}

\newpage
\appendix

\section{Video Processing}
\par To roughly check the rationality of the maximum interframe difference algorithm and the threshold, we conducted a manual review of the 69 key frames extracted from a sample video. Upon analysis, 29 key frames were found to be duplicates, with changes limited to instructor gesture and cursor movements, window scaling, and shifting. Additionally, we observed that the server discarded 9 out of 41 slides, deeming them redundant. The content examination revealed that the discarded slides only had minor variations from their adjacent ones such as non-essential textual elements or color variations. This exclusion did not impede the subsequent processes of content recognition and knowledge extraction, as the key information was preserved in the remaining key frames.
\begin{figure}[hb]
    \centering
    \includegraphics[width=\linewidth]{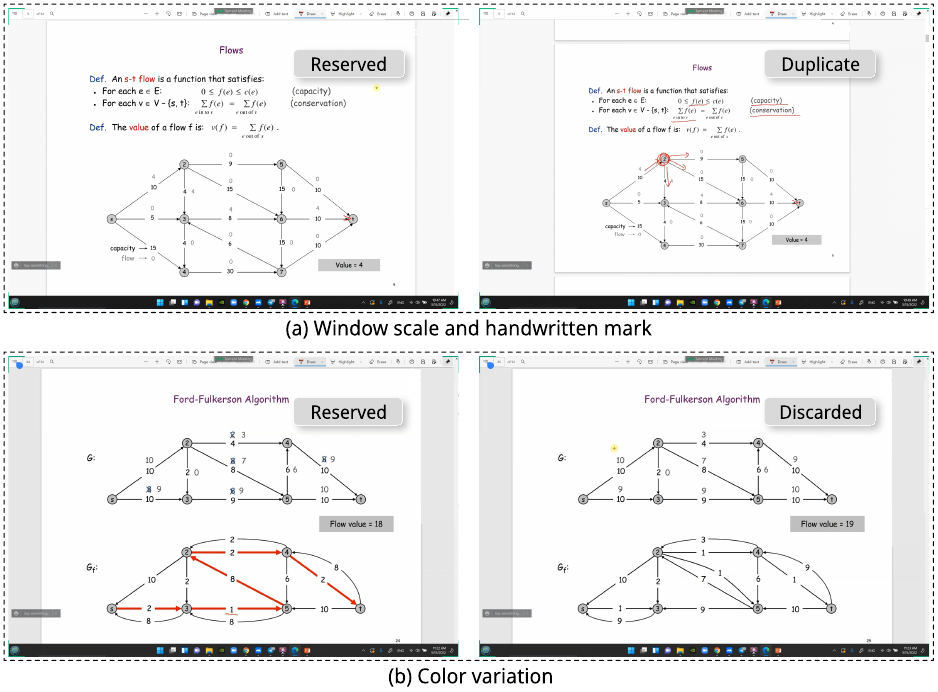}
    \caption{Illustrations of abnormal key frame extraction outcomes. (a) Key frame duplication: the server retains two instances of slide \#5 as key frames due to significant differences in window scaling and the presence of handwritten annotations. (b) Key frame discard: slide \#25 was discarded as a key frame candidate due to limited edge color variations.}
    \label{fig:FrameCompare}
\end{figure}

\section{Students' Preferences for Assessing Their Knowledge Mastery.}
\label{App:SelfevaPrefer}
\begin{figure}[ht]
  \centering
  \includegraphics[width=\linewidth]{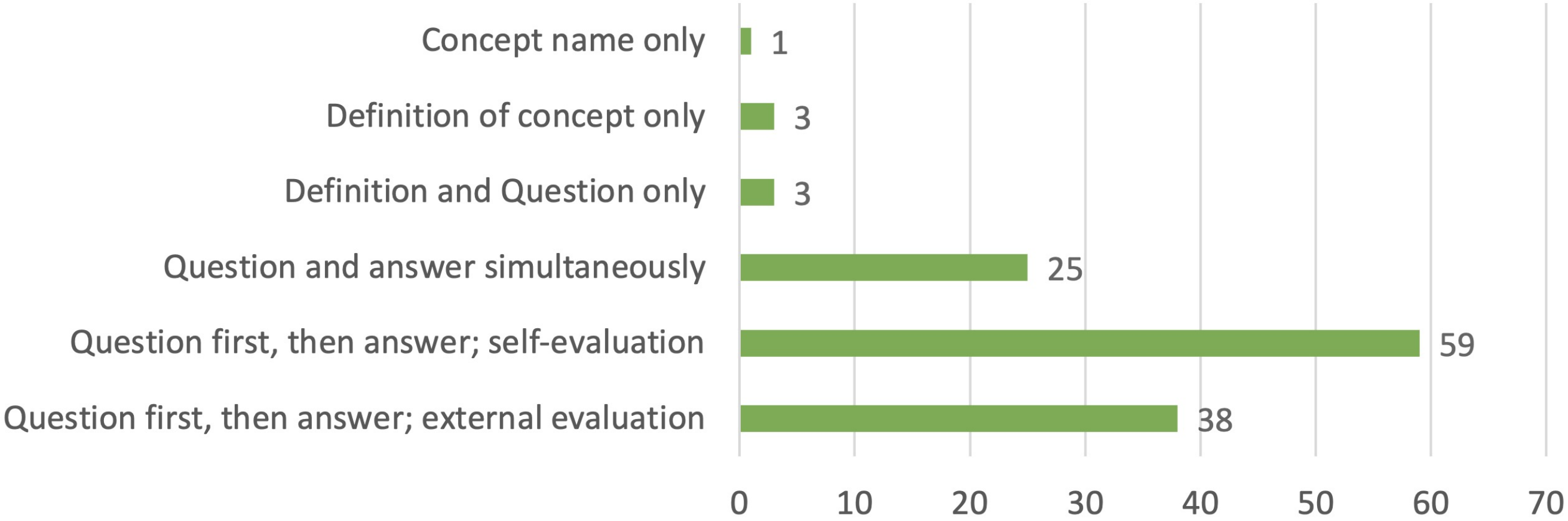}
  \caption{Question Description: If you are required to self-assess and report your knowledge mastery, which method do you think is more reasonable?}
\end{figure}

\end{document}